# Large Gain Degradation of Reflective Intelligent Surfaces in Realistic Environments

Dmitry Chizhik, *Fellow, IEEE*, Jinfeng Du, *Senior Member*

***Abstract*— Reflective Intelligent Surfaces (RIS) are considered promising in improving coverage in Non-Line of Sight (NLOS) wireless links, especially at mm wave or higher frequency bands. Coverage provided by RIS is here compared to coverage from such ambient propagation mechanisms as scattering from street poles (e.g. lampposts), and corner diffraction. A simple formula for RIS gain degradation due to channel angle spread is derived. It is found an ideal 0.3 m × 0.3 m RIS at 28 GHz promises to deliver only about 5 dB more power at 200 m around an urban street corner than the ambient scatter already there. Consideration of angle spread brings about some 14 dB drop in RIS power, bringing it well below ambient mechanisms. A 1 m × 1 m RIS at 28 GHz, offers under 2 dB advantage over ambient scatter after including the 25 dB gain degradation due to angle spread. This raises questions about usefulness of RIS-assisted coverage extension in realistic environments.**

## I. Introduction

Providing coverage for high-rate links in mm wave bands is challenging, especially in Non-Line of Sight (NLOS) scenarios. Reflective Intelligent Surfaces (RIS) have been proposed [1]-[5] as an economical way to improve coverage without deploying additional base stations. Typically, RIS is a two-dimensional array of small antenna elements, often idealized as being able to co-phase signal components scattered from its elements through adjusting the individual load impedance of each element. Determination of the necessary element phase shifts is challenging in dynamic environment such as cellular networks, particularly since RIS is usually billed as an inexpensive device, lacking the RF chains and receiver processing present in usual communication equipment. One simplifying assumption is that RIS would do a directional search to determine which incident direction is best coupled to which scattered direction, thus limiting the search space. This may appear effective in free space-like environment where clear LOS link exists with few to no multipaths between the transmitter and the RIS as well as between the RIS and the receiver. However, since propagation through real channels often involves multipath with different directions, such a solution is likely to suffer a degradation due to angle spread, which is assessed here.

Dmitry Chizhik and Jinfeng Du are with Nokia Bell Labs, Murray Hill, NJ 07974, USA (dmitry.chizhik@nokia-bell-labs.com, jinfeng.du@nokia-bell-labs.com,). (*Corresponding author: Dmitry Chizhik*).

In addition, RIS performance needs to be assessed not in absolute terms of the transmitter-RIS-receiver link itself but rather relative to other ambient mechanisms, which are already present in the channel, such as scattering and diffraction, which deliver signal in NLOS conditions. These mechanisms are what provides coverage to NLOS users in existing communication systems. Innovations such as RIS hope to improve upon that.

An ideal RIS was compared to ambient mechanisms (mostly street pole scatter) for around-the-corner propagation [10]. It was found that to deliver gain, RIS needed to be larger than 0.3 m × 0.3 m at 28 GHz, already some 10 times the size of a typical commercial base station phased array at 28 GHz. Recently, it was found that a RIS needs to be wider than 1 m to deliver signal power comparable to ambient mechanisms in indoor environments [14].

In [18] it was reported that angular spread leads to saturation of RIS coherent gain as number of RIS antenna elements increases. This is a manifestation of the general gain behavior of antenna arrays, e.g. Fig 7 in [19]. Considered environment had angle spread in the channel between the user terminal and RIS, while the channel between base station and RIS was assumed free of scattering, i.e. without angle spread.

In present work, we account for scattering-induced angle spread on both base -RIS and RIS-terminal segments of the composite base-RIS-terminal channel. Scattering on both segments is found to lead to up to 4× further reduction of power delivered by RIS, as compared to the case of scattering on only one segment. Total power (including path loss and directional gain) delivered by RIS and degraded by angle spread are here compared against power delivered by the ambient environment, mostly street poles, to determine net gain in power provided by RIS as opposed to relying on the environment.

Main contributions of current work include:

- Simple formula for cylindrical street pole scatter is found to accurately reproduce observed power in 28 GHz around the corner measurements. It is found to dominate over corner diffraction at higher frequencies.
- Formulation allowing assessment of the impact of angle spread on power delivered by a realistic, re-directing RIS is derived as a simple formula, parametrized on channel angle spread values in the base-RIS and RIS-terminal segments. In the presence of angle spread, power delivered by an electrically large RIS is found to scale with only RIS area, as opposed to a square of the RIS area, robbing the RIS of much of its intended benefit.

Angle spreads recommended by 3GPP standards lead to a severe degradation of simple RIS performance, raising questions about viability of coverage extensions via such RIS.

## II. AMBIENT AROUND-THE-CORNER PROPAGATION

Around-the-corner propagation mechanisms are now considered: diffraction around four building corners, scatter from four cylindrical street poles and a RIS scattered component. Some of these signal paths are illustrated in Fig.1. Scattering from rough building façade has also been considered in [10], but is omitted here for simplicity as it is weaker than other mechanisms. The street pole scatter in [10] treated the poles as octagonal with 8 facets. In this work the poles are treated as cylinders, which is far simpler yet produces a numerically close result to octagonal poles.

### A. Diffraction around corner

Diffraction at each of 4 building corners at the intersection is treated using Geometric Theory of Diffraction (GTD) formulas for wedges [6], with receive power

$$P_{\text{diffracted}} = \frac{P_{\text{T}} G_{\text{T}} G_{\text{R}} \lambda^2 |D|^2}{16\pi^2 r r' (r + r')} \tag{1}$$

where $r'$ and $r$ are the pre and post corner distances, respectively, and $G_{\text{T}}$ and $G_{\text{R}}$ are the transmit/receive antenna gains and transmit power $P_{\text{T}}$.

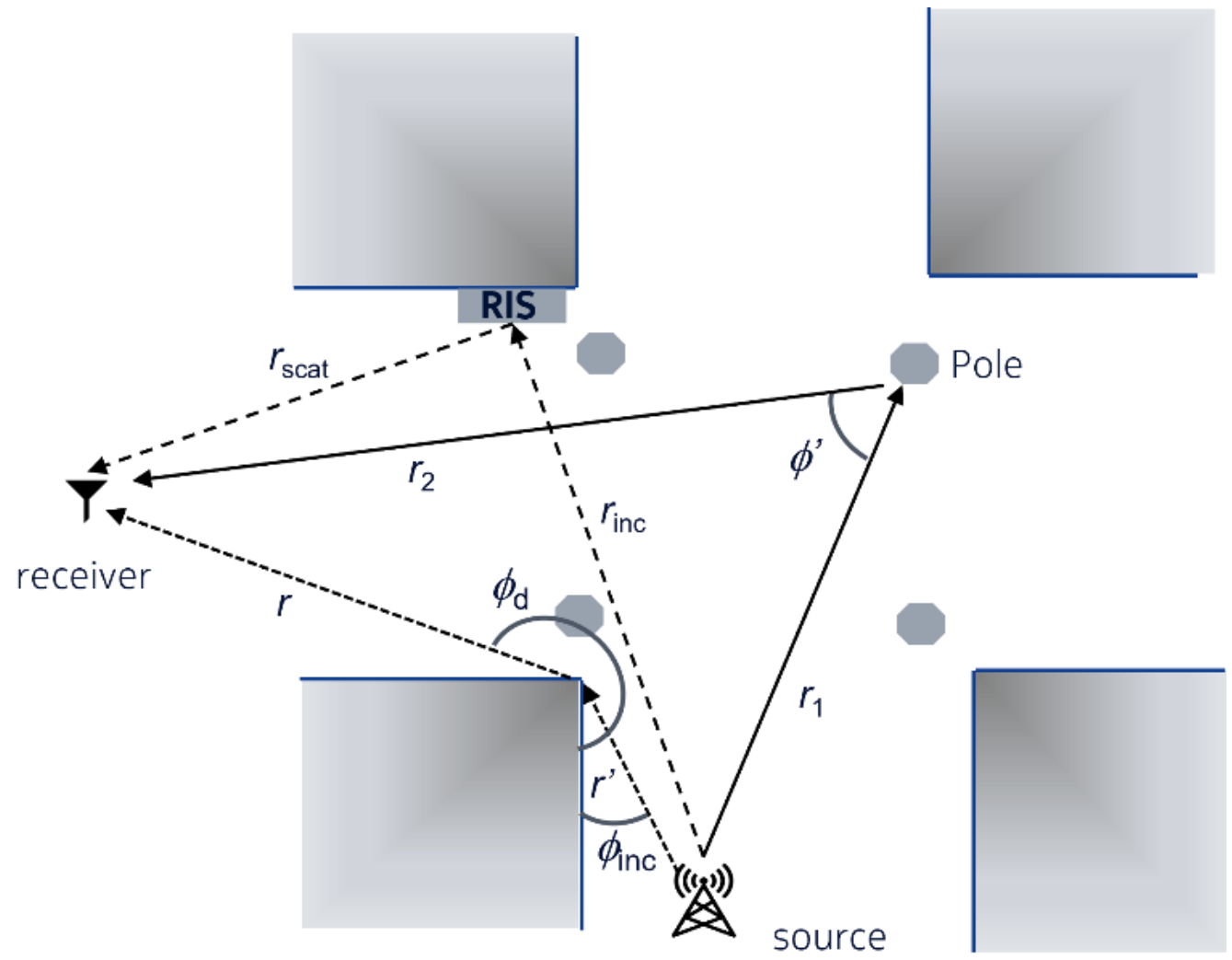

Fig. 1 Scattering mechanisms in around-the-corner propagation: corner diffraction, street pole scatter and RIS scatter.

The GTD diffraction coefficient D is dependent on the incident angle $\phi_{\text{inc}}$, and the diffracted angle $\phi_{\text{d}}$ through by the formula:

$$D \approx \frac{e^{i\pi/4} \sin(\pi/n)}{n\sqrt{2\pi k}} \times \left[ \frac{1}{\cos(\pi/n) - \cos\left(\frac{\phi_{\text{inc}} - \phi_{\text{d}}}{n}\right)} \pm \frac{1}{\cos(\pi/n) - \cos\left(\frac{\phi_{\text{inc}} + \phi_{\text{d}}}{n}\right)} \right] \tag{2}$$

Where $n$=3/2 for the building corner shaped as a right-angle wedge. The upper/lower sign is for hard/soft polarization, respectively. Wavenumber $k$ is related to wavelength $\lambda$ by $k = 2\pi/\lambda$. As considered geometry has both transmitter and receiver far away from shadow boundaries, simpler GTD formula (2) is chosen over the Uniform Theory of Diffraction (UTD) refinement [6].

### B. Street pole scatter

Here we consider scattering from 4 cylindrical poles, illustrated in Fig. 1 as a mechanism for around-the-corner propagation. Receive power due to bi-static scattering from an infinitely long pole of radius $a$ is given by

$$P_{\text{pole}} = \frac{P_{\text{T}} G_{\text{T}} G_{\text{R}} \lambda^2 \sigma_{\text{pole}} e^{-\kappa(r_1 + r_2)}}{2\pi (4\pi)^2 r_1 r_2 (r_1 + r_2)}$$

where the bi-static scattering width $\sigma_{\text{pole}}$ of a perfectly conducting pole of radius $a$ is given as a function of scattering angle $\phi'$ (Fig. 1) by (4.1-37) in [11]:

$$\sigma_{\text{pole}} = \pi a \cos(\phi'/2) \tag{4}$$

applicable in the high frequency limit $ka > 20$, valid for frequencies above 8 GHz for a (standard New York City) pole of 0.12 m radius. Formula (4) applies for $|\pi - \phi'| > (ka)^{-1/3}$ i.e. for scattering angles sufficiently away from forward scatter direction (i.e., $\phi' = \pi$), satisfied for the around the corner case of $\phi' \approx \pi/2$ for frequencies above 1.5 GHz.

## III. RIS CONTRIBUTION

### A. Zero angle spread

A RIS of total area $A_{\text{RIS}}$ is considered mounted on a building wall facing the source, near the intersection, Fig. 1. Assuming, for non-scattering case, unimpeded Line of Sight (LOS) between intersection and both transmitter and receiver, RIS-scattered field at the receiver is given by the scattering integral [6] for omnidirectional antennas:

$$U_{\text{RIS}} \approx 2ik \cos\theta_{\text{inc}} \iint_{A_{\text{RIS}}} dA' \, U_{\text{scat}}(y',z') U_{\text{inc}}(y',z') e^{i\phi_{\text{RIS}}(y',z')} \tag{5}$$

where $U_{\text{inc}} = \frac{e^{ikr_{\text{inc}}}}{4\pi r_{\text{inc}}}$ and $U_{\text{scat}} = \frac{e^{ikr_{\text{scat}}}}{4\pi r_{\text{scat}}}$ represent propagation of incident and RIS-scattered from source to RIS from RIS to receiver, respectively. The incidence angle $\theta_{\text{inc}}$ is relative to normal to RIS surface. Distances $r_{\text{inc}}$ and $r_{\text{scat}}$ vary over the surface of the RIS. Phase variation $\phi_{\text{RIS}}$ represents the phase shift imposed by the RIS as it adapts to the channel to maximize RIS scattered power. Expanding field phase variation over the RIS surface to 1st order gives:

$$U_{\text{RIS}} \approx 2ik \left\{ \frac{e^{ikR_{\text{scat}}}}{4\pi R_{\text{scat}}} \frac{e^{ikR_{\text{inc}}}}{4\pi R_{\text{inc}}} \right\} \cos\theta_{\text{inc}} \int_{-h/2}^{h/2} \int_{-w/2}^{w/2} dy' dz' e^{ik\boldsymbol{\rho}' \cdot (\hat{s} - \hat{o})} e^{i\phi_{\text{RIS}}(y',z')} \tag{6}$$

Where $\boldsymbol{\rho}' = (y', z')$, while $\hat{s}$ and $\hat{o}$ are, respectively, unit vectors from transmitter to RIS center and RIS center to receiver with $R_{\text{inc}}$ and $R_{\text{scat}}$ the corresponding distances . The integral limits in (6) are for a $w \times h$ rectangular RIS. When RIS has perfect channel knowledge, adaptive phase $\phi_{\text{RIS}}$ can be adjusted to cancel out phase variation given by $e^{ik\boldsymbol{\rho}' \cdot (\hat{s}-\hat{o})}$, reducing the integral in (6) to RIS geometric area $A_{\text{RIS}}$.

Spatial distribution of the real part of $e^{ik\boldsymbol{\rho}' \cdot (\hat{s}-\hat{o})}$ is illustrated in Fig. 2, for incident elevation and azimuth angles of 10° and scattered elevation and azimuth angles of -10°. The periodic spatial pattern is a consequence of both illuminating and scattered wavefronts being single plane waves. It is the task of the adaptive RIS to estimate $\hat{s}-\hat{o}$ to maximize the integral (6) through $\phi_{\text{RIS}}(y', z') = -k\boldsymbol{\rho}' \cdot (\hat{s}-\hat{o})$. The estimation is effectively of 2 parameters: relative azimuth and elevation angles.

Under such matched conditions and with transmit/receive antenna gains $G_{\text{T}}$ and $G_{\text{R}}$, the RIS contribution to received power

$$P_{\text{RIS}} = \lambda^2 P_{\text{T}} G_{\text{T}} G_{\text{R}} \left| U_{\text{RIS}} \right|^2 \tag{7}$$

is given by [7] (adding a factor of $\exp(-\kappa R_{\text{scat}} - \kappa R_{\text{inc}})$ to account for signal power loss due to absorption in channel):

$$P_{\text{RIS}} \approx \frac{P_{\text{T}} G_{\text{T}} G_{\text{R}} A_{\text{RIS}}^2 \cos^2\theta_{\text{inc}} \exp(-\kappa R_{\text{scat}} - \kappa R_{\text{inc}})}{(4\pi)^2 R_{\text{scat}}^2 R_{\text{inc}}^2} \tag{8}$$

where $R_{\text{scat}}$ and $R_{\text{inc}}$ are the distances between the RIS center and receiver and transmitter, respectively. Neglected are limitations imposed by circuits which can only approximately deliver an adjustable phase.

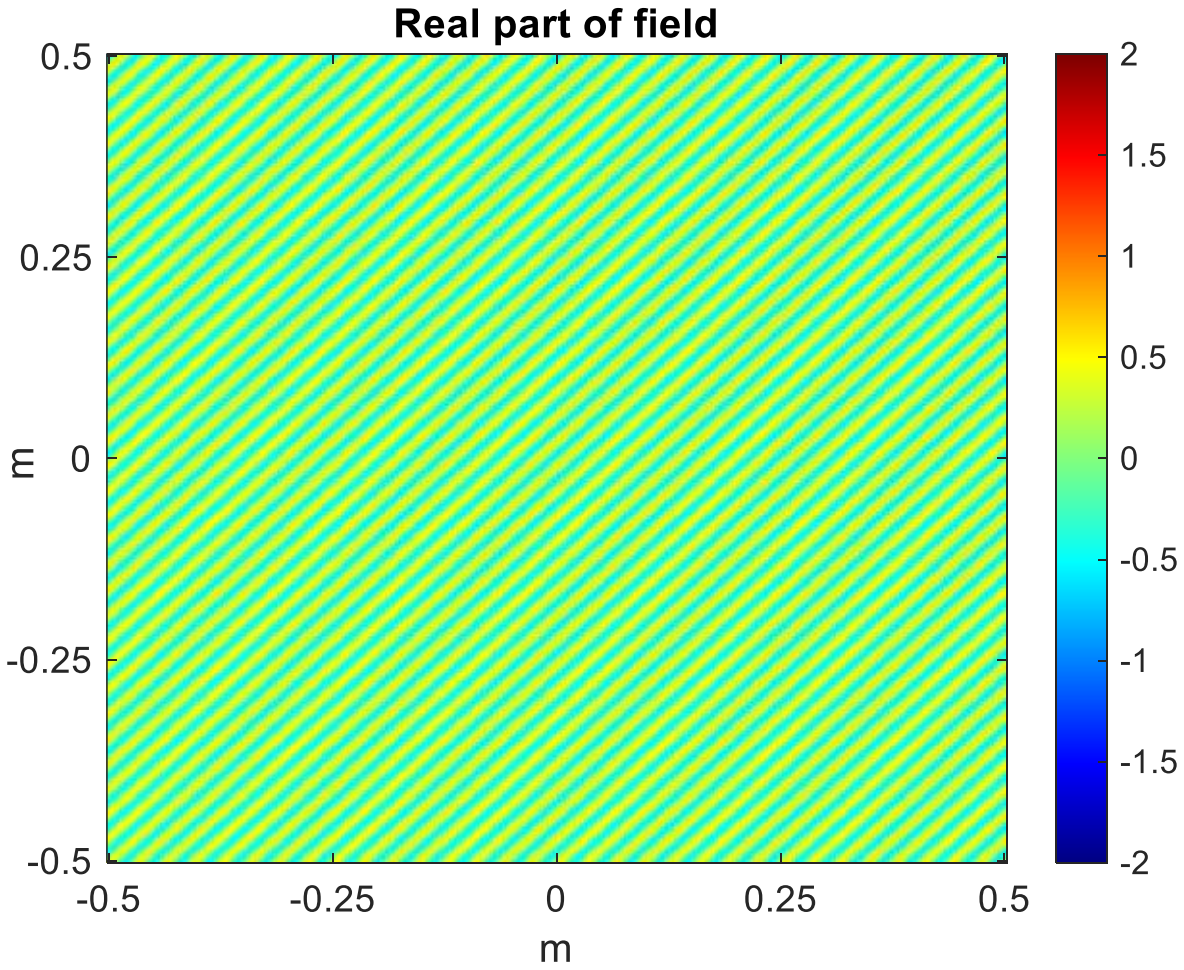

Fig. 2 Real part of $e^{ik\boldsymbol{\rho}' \cdot (\hat{s}-\hat{o})}$ over 1×1 m aperture for incident elevation and azimuth angles of 10° and scattered elevation and azimuth angles of -10°.

Notably, for given transmit and receive antenna gains, power delivered by ideal RIS (8) in these unobstructed conditions is independent of frequency. This is unlike non-adaptive channels, where receive power typically depends on frequency, for example scaling as $\lambda^2$ in free space conditions. Coherent gain across the RIS delivered by adapting RIS phase distribution is what is responsible for $A_{\text{RIS}}^2$ in (8) as opposed to $A_{\text{RIS}}$ expected from a non-adaptive (rough) scatterer of the same area.

### *B. Finite angle spread*

Propagation through environments that contain scattering objects, including trees, lampposts, vehicles and people, will reduce received power as well as introduce angle spread. Additional path loss due to scattering may be accounted for by including street-specific vegetation [12] or heuristically through using exponential absorption [13] factor scaling received power $\exp(-\kappa R_{\text{inc}} - \kappa R_{\text{scat}})$, where $\kappa$=0.005 Np/m, corresponding to 0.02 dB/m.

To represent angle spread, fields for Tx-RIS and RIS-Rx segments of the signal path may be written as nominal (unspread) fields scaled by a randomly distributed "de-coherence" factors $\xi_{\text{inc}}(x, y)$ and $\xi_{\text{scat}}(x, y)$:

$$\begin{aligned} U'_{\text{inc}} &= U_{\text{inc}} \xi_{\text{inc}}(y, z) \\ U'_{\text{scat}} &= U_{\text{scat}} \xi_{\text{scat}}(y, z) \end{aligned} \tag{9}$$

It is now assumed that scattering leads to Rayleigh fading which corresponds to zero-mean, unit variance complex Gaussian fading:

$$\xi_{\text{inc}}(y, z) \sim CN(0,1),\ \xi_{\text{scat}}(y, z) \sim CN(0,1) \tag{10}$$

With spatial correlation

$$\rho(y_{\text{d}}, z_{\text{d}}) \triangleq \left\langle \xi(y, z) \xi^*(y + y_{\text{d}}, z + z_{\text{d}}) \right\rangle \tag{11}$$

Here assumed to have Gaussian form:

$$\rho(y_{\text{d}}, z_{\text{d}}) = \exp\left(-k^2 y_{\text{d}}^2 \phi_{\text{rms}}^2 / 2 - k^2 z_{\text{d}}^2 \theta_{\text{rms}}^2 / 2\right) \tag{12}$$

Parametrized by channel azimuth spread $\phi_{\text{rms}}$ and elevation spread $\theta_{\text{rms}}$, both expressed in radians. For spatially stationary processes, the spatial correlation is the inverse Fourier transform of the plane wave power spectrum. Replacing $U_{\text{inc}}$ and $U_{\text{scat}}$ fields in (5) by corresponding primed fields from (9), leads to

$$U_{\text{RIS}} \approx 2ik \left\{ \frac{e^{ikR_{\text{scat}}}}{4\pi R_{\text{scat}}} \frac{e^{ikR_{\text{inc}}}}{4\pi R_{\text{inc}}} \right\} \cos\theta_{\text{inc}}$$
$$\times \int_{-h/2}^{h/2} \int_{-w/2}^{w/2} dy' dz' \xi_{\text{inc}}(y', z') \xi_{\text{scat}}(y', z') \exp\left(ik\boldsymbol{\rho}' \cdot (\hat{s}-\hat{o}) + i\phi_{\text{RIS}}(y', z')\right) \tag{13}$$

Which now includes the scattering of fields from transmitter to RIS and from RIS to receiver.

The total contribution delivered by RIS as represented by the integral in (13) is thus a "summation" of the field shown in Fig. 3 over the area of the RIS. It may be observed that when RIS area exceeds the characteristic spatial scales of the random field, different parts of the RIS deliver signal with different phases. Now a RIS with perfect knowledge of the channel could still adjust its phase $\phi_{\text{RIS}}(y',z')$ across the RIS aperture to render the integrand in (13) non-negative, leading to strong RIS contribution, similar to (8). A simulated instantiation of the real part of $\xi_{\text{inc}}(y,z)\xi_{\text{scat}}(y,z)$, defined by (10) and (12), is shown for illustration in Fig. 3. The imaginary part is statistically similar. Such field phase distribution across the RIS would need to be estimated in situ to deliver a fully coherent RIS gain. This is contrast to a 2-parameter (relative azimuth and elevation angles) estimation needed to compensate simple phase variation in (6) in the absence of angle spread, illustrated in Fig. 2. In addition, for wideband signals used in modern systems, a potentially different phase distribution would need to be estimated for every coherence bandwidth. It is assumed in this work that a practical system using a RIS would not have the ability to make corresponding estimates, allowing only estimates of general direction to transmitter and receiver to be available allowing RIS to compensate for $e^{ik\boldsymbol{\rho}'\cdot(\hat{s}-\hat{o})}$ only in (13). Under this assumption, (13) becomes

$$U_{\text{RIS}} \approx 2ik\left\{\frac{e^{ikR_{\text{scat}}}}{4\pi R_{\text{scat}}}\frac{e^{ikR_{\text{inc}}}}{4\pi R_{\text{inc}}}\right\}\cos\theta_{\text{inc}} \times\int_{-h/2}^{h/2}\int_{-w/2}^{w/2} dy'dz'\xi_{\text{inc}}(y',z')\xi_{\text{scat}}(y',z') \tag{14}$$

where the integrand has uncompensated spatial phase variation, leading to partial mutual cancellation. The impact of such cancellation on average power delivered by RIS is evaluated below.

The average received power from RIS

$$P_{\text{RIS}} = \lambda^2 P_{\text{T}} G_{\text{T}} G_{\text{R}} \left\langle \left|U_{\text{RIS}}\right|^2 \right\rangle \tag{15}$$

is then given by (adding a factor of $e^{-\kappa(R_{\text{scat}}+R_{\text{inc}})}$ to account for signal power loss due to scatter):

$$\begin{aligned} P_{\text{RIS}} &\approx \frac{P_{\text{T}}G_{\text{T}}G_{\text{R}}\cos^2\theta_{\text{inc}}e^{-\kappa(R_{\text{scat}}+R_{\text{inc}})}}{(4\pi)^2 R_{\text{scat}}^2 R_{\text{inc}}^2}\iint dAdA_{\text{d}}\rho_{\text{inc}}(y_{\text{d}},z_{\text{d}})\rho_{\text{scat}}(y_{\text{d}},z_{\text{d}}) \\ &= \frac{P_{\text{T}}G_{\text{T}}G_{\text{R}}\cos^2\theta_{\text{inc}}A_{\text{RIS}}e^{-\kappa(R_{\text{scat}}+R_{\text{inc}})}}{(4\pi)^2 R_{\text{scat}}^2 R_{\text{inc}}^2}\iint dy_d dz_d\rho_{\text{inc}}(y_{\text{d}},z_{\text{d}})\rho_{\text{scat}}(y_{\text{d}},z_{\text{d}}) \\ &= \frac{P_{\text{T}}G_{\text{T}}G_{\text{R}}\cos^2\theta_{\text{inc}}A_{\text{RIS}}A_{\text{eff}}e^{-\kappa(R_{\text{scat}}+R_{\text{inc}})}}{(4\pi)^2 R_{\text{scat}}^2 R_{\text{inc}}^2} \end{aligned} \tag{16}$$

where $dA_{\text{d}} = dy_d dz_d$ and the "effective coherent RIS area" $A_{\text{eff}}$ is defined as

$$\begin{aligned} A_{\text{eff}} &\triangleq \iint_{A_{\text{RIS}}} dz_{\text{d}}dy_{\text{d}}\rho^2(y_{\text{d}},z_{\text{d}}) \\ &\approx \min(w_{\text{coh}},w)\times\min(h_{\text{coh}},h) \end{aligned} \tag{17}$$

for a rectangular RIS dimensioned as $w\times h$ and correlation $\rho(y_{\text{d}},z_{\text{d}})$ separable as in (12). The field coherence width $w_{\text{coh}}$ and coherence height $h_{\text{coh}}$ depend on the wavelength $\lambda$ as well as azimuth spread $\phi_{\text{rms}}$ and elevation spread, $\theta_{\text{rms}}$. For Gaussian correlation (12):

$$w_{\text{coh}} = \frac{\lambda}{2\sqrt{\pi}\phi_{\text{rms}}}, h_{\text{coh}} = \frac{\lambda}{2\sqrt{\pi}\theta_{\text{rms}}} \tag{18}$$

The result (16) can be interpreted as follows: under angular spread, fields $\xi_{\text{inc}}(y',z')\xi_{\text{scat}}(y',z')$ on RIS within a small coherent region of dimension $w_{\text{coh}}\times h_{\text{coh}}$ have (roughly) the same phase, and therefore would deliver coherent combining over an area $A_{\text{eff}} = w_{\text{coh}}\times h_{\text{coh}}$, leading to a power gain proportional to $A_{\text{eff}}^2$. For a larger RIS of area $A_{\text{RIS}}$, there are about $A_{\text{RIS}}/A_{\text{eff}}$ of such coherent regions, each with its own phase, delivering a further non-coherent (power) combining gain proportional to $A_{\text{RIS}}/A_{\text{eff}}$. The two effects, together, deliver a total gain proportional to $\left(A_{\text{RIS}}/A_{\text{eff}}\right)\times A_{\text{eff}}^2 = A_{\text{RIS}}A_{\text{eff}}$, as seen in (16).

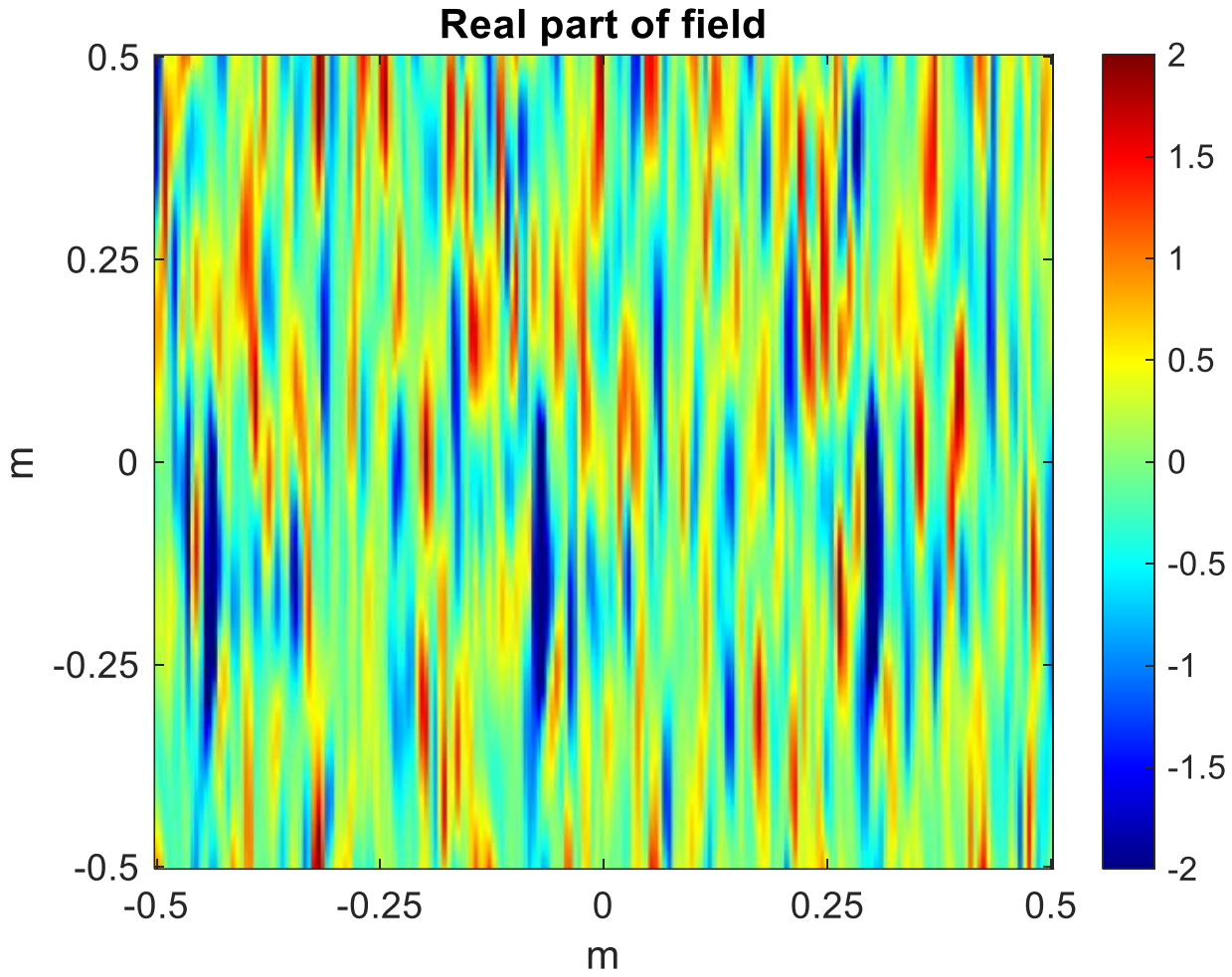

Fig. 3 Simulated instantiation of real part of $\xi_{\text{inc}}(y,z)\xi_{\text{scat}}(y,z)$ distribution over 1×1 m aperture for 14º RMS azimuth, 0.6º RMS elevation spreads.

The approximation (17) is used to avoid the use of error functions which would be the exact result of integrating a Gaussian correlation over finite intervals imposed by finite RIS area. The above definition includes the special case of coherent fields with rms angle spreads tending to zero, in which case $A_{\text{eff}} = A_{\text{RIS}}$, reducing (16) to (8). Evaluating (18) at 28 GHz $\phi_{\text{rms}} = 14^{\text{o}}, \theta_{\text{rms}} = 0.6^{\text{o}}$ (typical[1] angular spread for AOD, ZOD recommended by 3GPP Umi channels in LOS [9]), coherence scales are found to be $w_{\text{coh}}$=0.01 m and $h_{\text{coh}}$= 0.28 m.

For a 1×1 m RIS (thus $A_{\text{RIS}}$=1 m$^2$), the effective RIS area $A_{\text{eff}}$ ~ 0.0035 m$^2$. Angle spread thus produces a $10\log_{10}(0.0035)\sim 25$ dB reduction in RIS power (16) as compared to what might be expected using (8) from a 1 m$^2$ RIS in the absence of angle spread.

[1] Median RMS angular spread in UMi LOS [9] for azimuth is a function of carrier frequency, ranging from 13.7º to 14.6º for frequency of 7 to 28 GHz. For elevation, the median RMS spread decreases with distance [9], with a minimum spread of 0.6º. We take 14º RMS azimuth and 0.6º RMS elevation spreads as reference for evaluation.

The case of RIS being in LOS of the base station, without scatter, (as in [18]) corresponds to setting $\rho_{\text{inc}}(y'_d, z'_d) = 1$ in (16), leading to doubling of $w_{\text{coh}}$ and $h_{\text{coh}}$ in (18), thus increasing coherent RIS area $A_{\text{eff}}$ (17) by up to 4×.

## IV. AMBIENT AND RIS SCATTERED PATH GAIN COMPONENTS

In Fig. 4, path gain contributions at 28 GHz from RIS, ambient scatter as well as the various propagation mechanisms described above are plotted vs. distance between the intersection and the receiver, assuming the transmitter is 100 m from the intersection. Path gain is obtained from corresponding receive power by normalizing it by transmit power $P_T$ and transmit and receive antenna gains $G_T$ and $G_R$. Also shown is the empirical fit to around-the-corner path gain measurements [7] collected in New York City, needed to validate theoretical predictions of the ambient scatter.

Various ambient scattered contributions were modeled as described above, with the additional absorption loss of 0.02 dB/m to fit LOS observations in [7], attributed to the scattering from vegetation and other street clutter. It may be observed that scattering near the intersection from the 4 street poles represents the observed path gain within 2 dB. Corner diffraction is seen to be a much weaker mechanism as compared to both observed received power and pole scatter predictions. Corner diffracted power plotted in Fig. 4 is for hard polarization. Soft polarization diffraction would yield even weaker predicted power. It was found in [15] that 4-corner diffraction contribution is comparable at long distances to single pole scatter at 1.8 GHz. This changes substantially as the center frequency increases. Corner diffracted power is proportional (1) to $|D|^2 \sim 1/k$ (2), thus the diffraction excess loss scales as 1/frequency. In contrast, scattering width of a pole (4) is independent of frequency for $ka >> 1$. Thus, for higher frequencies considered here, pole scatter delivers higher power than corner diffraction. Furthermore, four poles present at most intersections in a city like Manhattan deliver 6 dB more power than a single pole. It was found experimentally that excess loss around intersections is effectively independent of frequency in the range 0.43 MHz to 4.86 GHz [16], while measured scatter from a pole was found to be substantially stronger at 60 GHz than corner diffraction in [17].

It is also seen that an ideal 0.3 m × 0.3 m RIS in the absence of angle spread can deliver about 14 dB more power at 20 m from the intersection than the ambient scatter already present. The gain reduces to merely 5 dB at 200 m from the intersection. Such a RIS is about 10× the area of a typical commercial base station array at 28 GHz. It would require the use of ~ 3600 elements at 28 GHz spaced 0.5 lambda apart.

When RMS angle spread values of 14° in azimuth and 0.6° in elevation are considered (as recommended in LOS Urban Microcell (Umi) channels in [9] through (16), (17),(18), it is seen in Fig. 4 that power delivered by a 0.3 m × 0.3 m RIS drops by 14 dB compared to the ideal RIS case, and well below ambient scatter at longer distances.

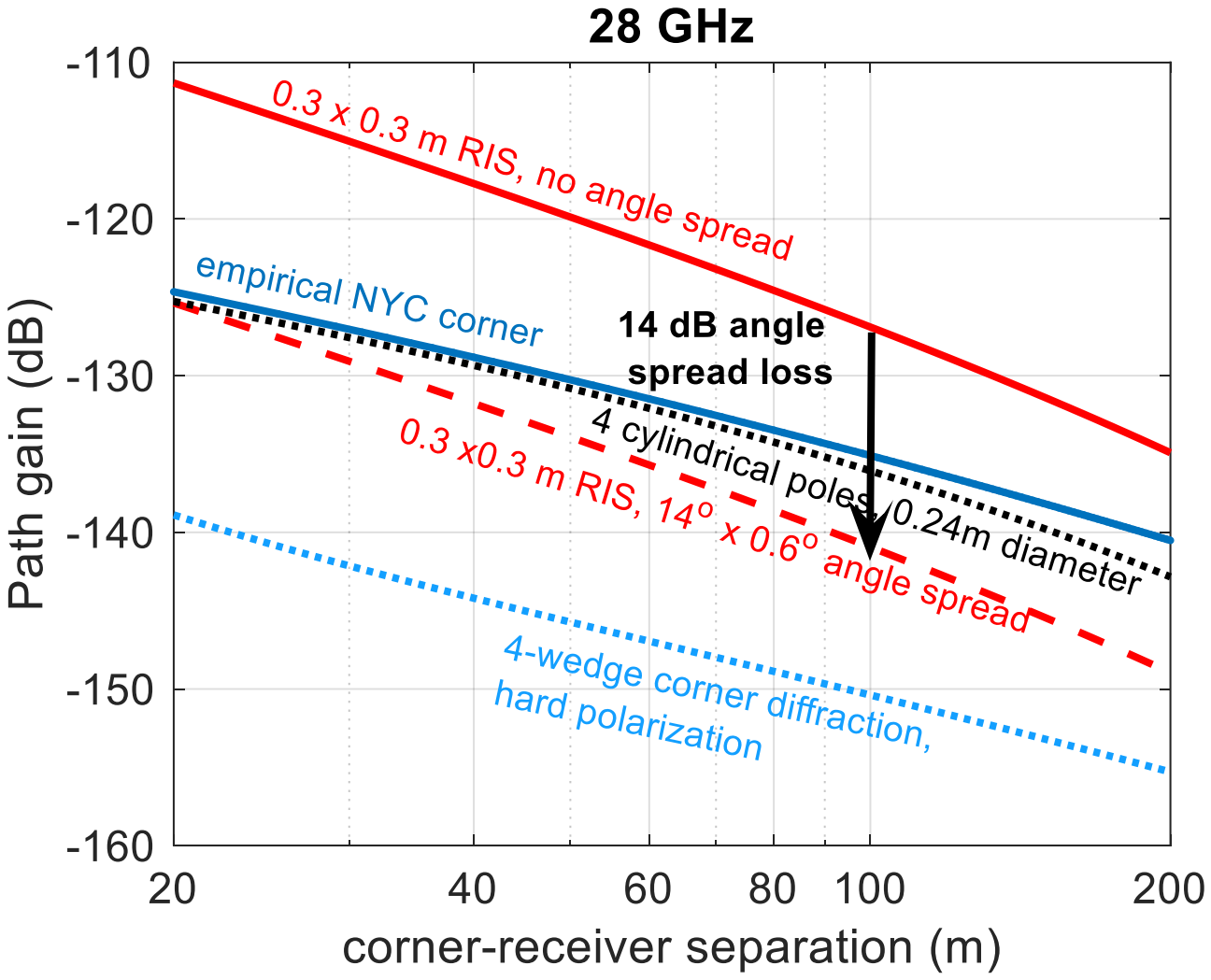

Fig. 4 Around-the-corner path gain from ambient and RIS-scattered components, source 100 m from the corner. RIS of size 0.3 m × 0.3 m is deployed at the corner.

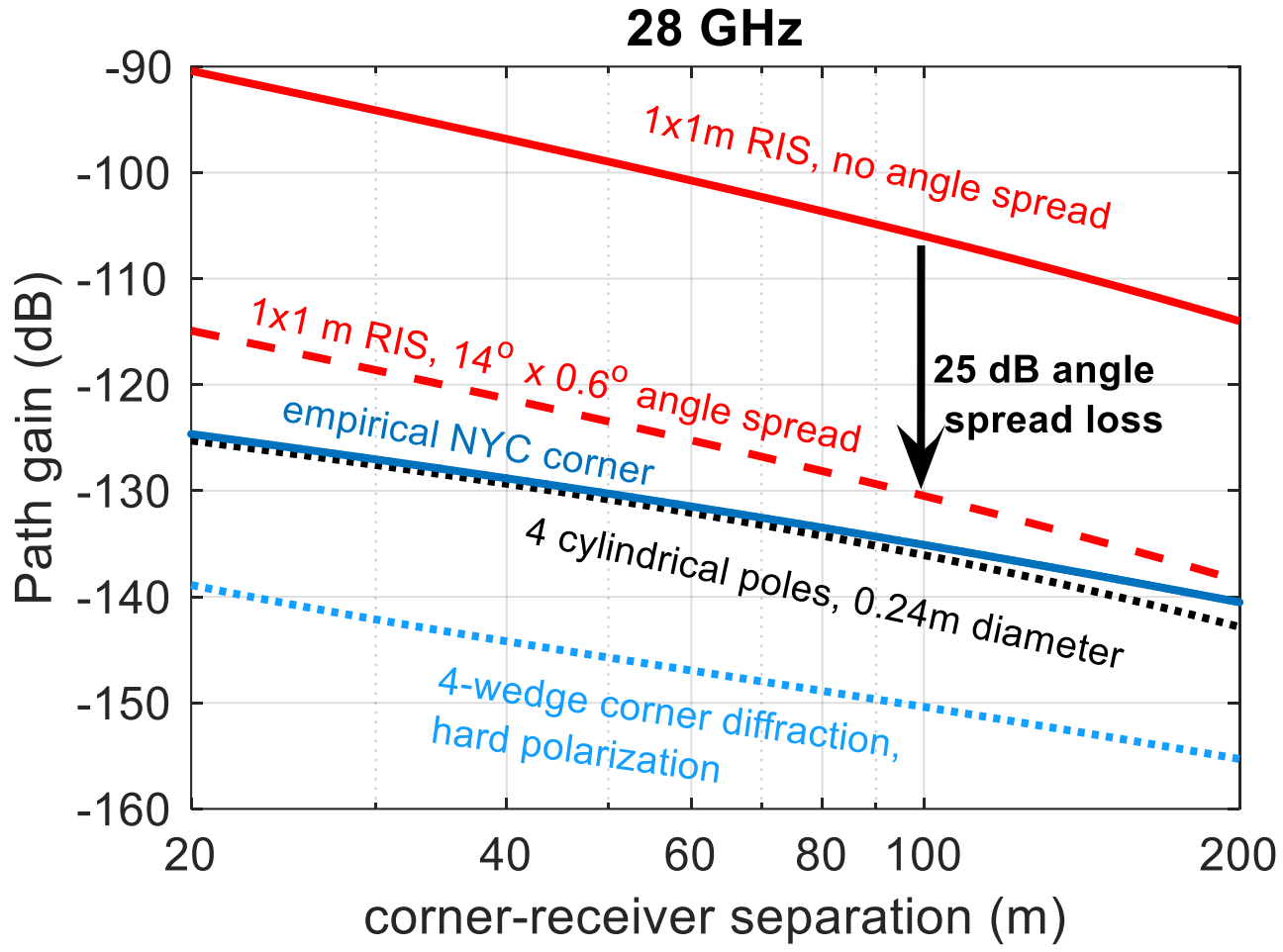

Fig. 5 Around-the-corner path gain at 28 GHz from ambient and RIS-scattered components, source 100 m from the corner. RIS of size 1 m × 1 m is deployed at the corner.

Power delivered by a more ambitious 1 m × 1 m RIS is seen in Fig. 5 to promise some 30 dB more power than ambient, in the absence of angle spread. When RMS angle spread values of 14° in azimuth and 0.6° in elevation are considered, as before, the estimated power delivered by RIS drops by some 25 dB, now promising under 2 dB gain over the ambient at 200 m beyond the street corner. Similar calculations at the 8 GHz center frequency of interest in the emerging 6G wireless communication band, lead to a similar conclusion: a 1 m × 1 m RIS under 3GPP recommended angle spread delivers under 3 dB signal gain over ambient scatter at 200 m, as seen in Fig. 6.

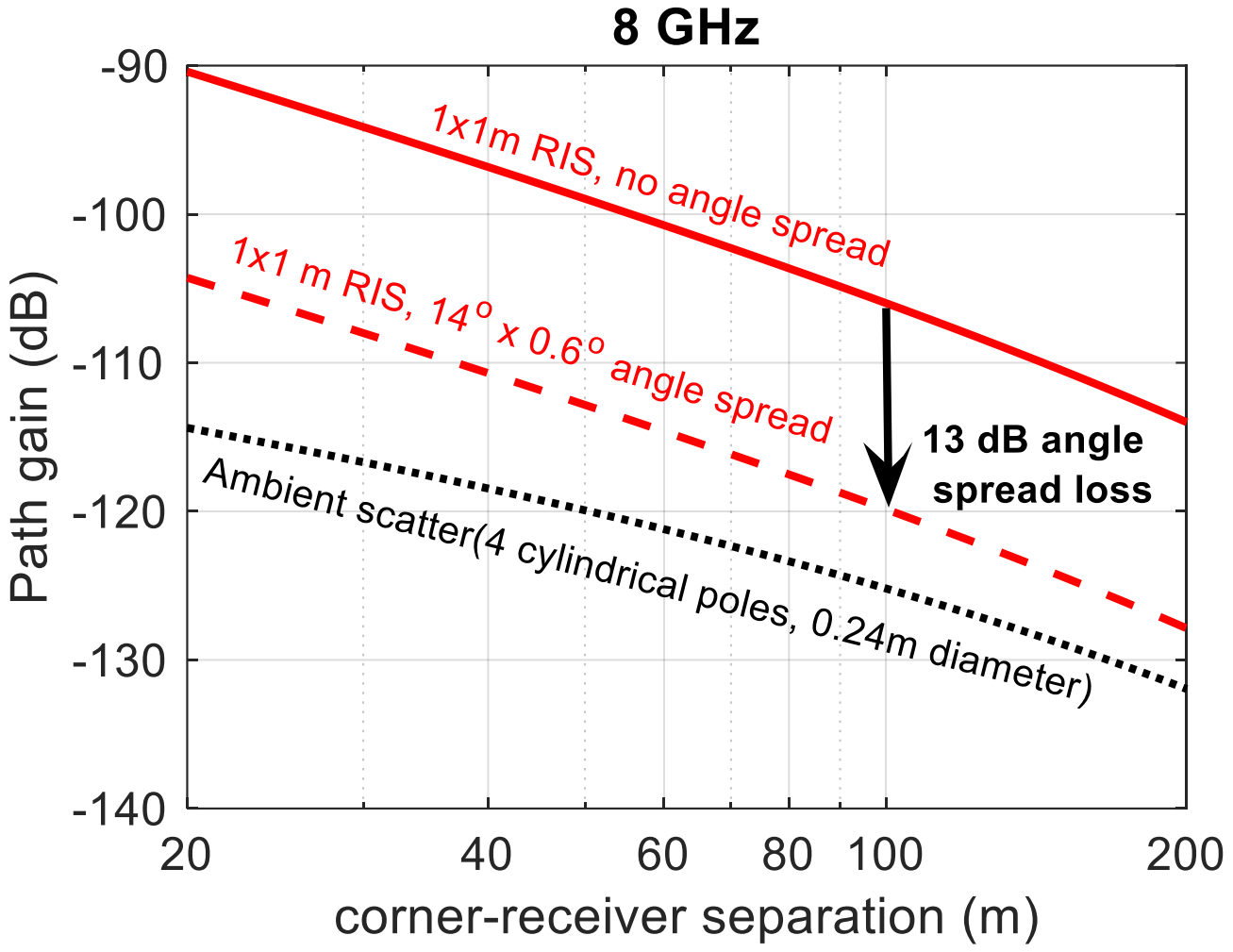

Fig. 6 Around-the-corner path gain at 8 GHz from ambient and RIS-scattered components, source 100 m from the corner. RIS of size 1 m × 1 m is deployed at the corner.

## V. Conclusion

Theoretical estimate of the ambient scatter present in around-the-corner propagation scenario (mostly pole scatter) is found to be within 2 dB of observations. This accurate model of the ambient environment is used as a reference to evaluate power gain delivered by the RIS.

An ideal 0.3 m × 0.3 m RIS (about 10 times the area of a typical base station array at 28 GHz, requiring phase control of about 3600 RIS elements of half wavelength dimension) is found to deliver only 5 dB more power at 200 m from the intersection than ambient channel scattering mechanisms. A much larger 1 m × 1m RIS (phase control of 40,000 elements separated by ½ wavelength), under realistic assumptions of RIS operation in presence of angular spreads, delivers under 2 dB more power than ambient at 200 m beyond the street corner.

As is already established, an ideal RIS, in the absence of channel angle spread, delivers signal whose power (8) scales as the square of RIS geometric area (8), while the number of RIS elements subject to phase control grows quadratically with carrier frequency. When channel has angular spread, the size of coherent region is proportional to wavelength squared, leading to a reduction in RIS power from the ideal that grows quadratically with increasing carrier frequency.

The degradation of RIS performance in these estimates is due to angle spread in the channel, which makes simple re-direction of the signal by a highly directional RIS inadequate. A direct remedy to this is well known in array communications: in angle spread channels, one needs an estimate of the channel at each array element to allow for coherent gain, beyond simple directional beamforming. Such estimates, however, need to be made practical when considering RIS, usually assumed to lack receiver structure to allow direct channel estimates.